\documentclass[twocolumn,doublespacing,showpacs,preprintnumbers,amsmath,amssymb]{revtex4}
\usepackage{amssymb}
\usepackage{txfonts}
\usepackage{graphicx}
\usepackage{subfigure}
\usepackage{dcolumn}
\usepackage{amsmath}
\usepackage{bm}
\usepackage{sidecap}

\begin{document}
\title{Continuous-variable quantum key distribution with
 Gaussian source noise}

\author{Yujie Shen}
\author{Xiang Peng}\thanks{E-mail: xiangpeng@pku.edu.cn}
\author{Jian Yang}
\author{Hong Guo }\thanks{E-mail: hongguo@pku.edu.cn}
\affiliation{CREAM Group, State Key Laboratory of  Advanced Optical
Communication Systems and Networks (Peking University) and Institute
of Quantum Electronics, School of Electronics Engineering and
Computer Science, Peking University, Beijing 100871, PR China}

\begin{abstract}
Source noise affects the security of continuous-variable quantum key
distribution (CV QKD), and is difficult to analyze. We propose a
model to characterize Gaussian source noise through introducing a
neutral party (Fred) who induces the noise with a general unitary
transformation. Without knowing Fred's exact state, we derive the
security bounds for both reverse and direct reconciliations and show
that the bound for reverse reconciliation is tight.
\end{abstract}

\maketitle

\section{INTRODUCTION}
Continuous-variable quantum key distribution helps two remote
parties (Alice and Bob) to establish a set of secret keys at high
speed \cite{Scarani_RMP_09}. Different from discrete-variable
protocols, in CV QKD Alice encodes information into the quadratures
of the optical field and Bob decodes it with high-efficiency and
high-speed homodyne detection
\cite{GG02,Grosshans_Nature_03,Weedbrook_PRL_04}. Besides the
experimental advantages and demonstrations, the security of CV QKD
is also studied theoretically. The coherent-state  CV-QKD protocol
with Gaussian modulation has been proved secure under the collective
attack
\cite{Grosshans_PRL_05,Navascues_PRL_05,Garcia_PRL_06,Navascues_PRL_06,Piradola_PRL_08},
and the fact that the security bounds for collective and coherent
attacks coincide asymptotically has been clarified using quantum De
Finetti theorem \cite{Scarani_RMP_09, Renner_PRL_09}. However, the
security of practical CV-QKD system has only been noticed recently
\cite{Lodewick_PRA_10, Lodewick_Ex_PRA_07}. It has been observed
that adding noise in the error-correction postprocessing may
increase the secret key rate \cite{Garcia_PRL_09}. Furthermore,
Filip \textit{et al}. noticed that the source noise in coherent
state preparation would undermine the key rate
\cite{Filip_PRA_08,Filip_PRA_10}. More recently, Weedbrook
\textit{et al.} has shown that direct reconciliation CV protocols is
more robust against this noise than reverse reconciliation protocols
\cite{Weedbrook_PRL_10}.

From a practical viewpoint, it is meaningful to consider the trusted
Gaussian source noise which is not controlled by the potential
eavesdropper Eve. To analyze this source-noise effect, it is
convenient to use the entanglement-based (EB) scheme to evaluate
CV-QKD security. Note that two requirements for the EB scheme should
be satisfied here. First, the EB scheme is kept equivalent to the
practical prepare and measure (PM) scheme \cite{Grosshans_QIC_03}.
Second, in the EB scheme the optimality of Gaussian attack is
guaranteed under the collective attack. From this viewpoint, a
three-mode entangled-state model has been proposed in
\cite{Shen_JPB_09} as a preliminary attempt. However, the security
bound in that paper is not tight, since in order to derive a
calculable bound the model assumes that the source noise is
untrusted, from which Eve is able to acquire extra information.
Another attempt \cite{Filip_PRA_10} used a beam-splitter model [Fig.
\ref{Fig-1}(a)], analogous to the realistic detector model
\cite{Lodewick_Ex_PRA_07}, to characterize the source noise.

In this paper, we propose a novel EB model to characterize the
Gaussian source noise. In this model a neutral party Fred introduces
Gaussian source noise through a general Gaussian transformation.
Without knowing Fred's exact state, a security bound can be derived
which is tighter than previous work \cite{Shen_JPB_09}. We also
analyze the performance of the beam-splitter model under situations
where source noise process includes either signal amplification or
attenuation, and make comparisons with our result.

\section{MODEL DESCRIPTION}
Before the explicit description of our model, the definition of the
covariance matrix of a quantum state is briefly reviewed. For an
N-mode quantum state, its covariance matrix $\gamma$ is defined by
\begin{equation}\label{cov}
\gamma_{ij}={\rm Tr}[\rho \{
(\hat{r}_{i}-d_{i}),(\hat{r}_{j}-d_{j})\}],
\end{equation}
where the operator vector is $\hat{r}=(\hat{x}_{1}, \hat{p}_{1},
\hat{x}_{2}, \hat{p}_{2},\ldots, \hat{x}_{N}, \hat{p}_{N})$ and the
displacement vector is $d_{i}={\rm Tr}(\rho \hat{r}_{i}) $ ($d\in
\mathbb{R}^{2N}$). $\hat{x_{i}}$ and $\hat{p_{i}}$ are the
quadratures of each optical field mode.

\begin{figure}
\centering
\includegraphics[width=8.5 cm]{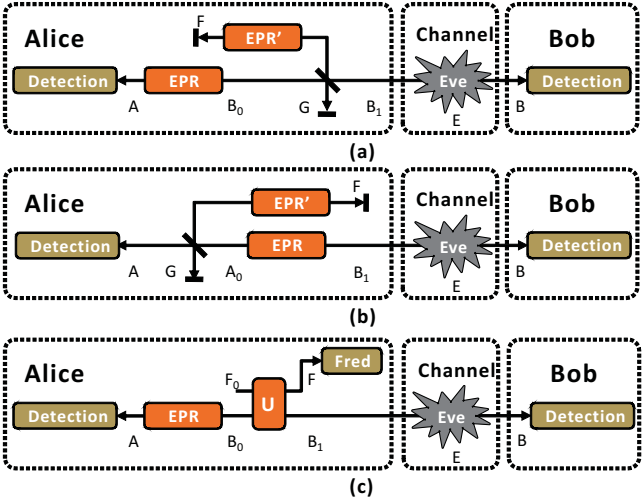}
\caption{(a) Beam-splitter model with signal attenuation. An extra
EPR state is presented and one mode of which is coupled with the
signal mode to introduce the source noise. This is similar to the
conventional detector model \cite{Lodewick_Ex_PRA_07}. (b)
Beam-splitter model with signal amplification. The extra EPR state
intervenes through coupling the mode sent to Alice and parameters
are adjusted so that the EB scheme is equivalent to the PM scheme.
(c) Our model. A neutral party Fred is presented who implements a
unitary Gaussian transformation to introduce the source noise to the
signal. The capital letters represent quantum states at each
position.}\label{Fig-1}
\end{figure}

The EB scheme of the beam-splitter model and our model is
illustrated in Fig. \ref{Fig-1}. In the beam-splitter model the
source is characterized by an EPR state held by Alice. Then an extra
EPR state interacts with either mode of the original one, depending
on whether the source noise process amplifies or attenuates the
signal, to introduce the noise. This is shown in Fig. 1(a) and Fig.
1(b).

Our model is demonstrated in Fig. 1(c) where the source is also
characterized by an EPR state. Alice obtains the data by measuring
one of mode $A$ and sends the other one $B_{0}$ to Bob as the
signal. We assume that Gaussian source noise is introduced by Fred
who implements a unitary Gaussian transformation over ${F_{0}}$ and
the signal ${B_{0}}$. The covariance matrix of the Gaussian state
$\rho_{FAB_{1}}$ describing Fred-Alice-Bob system after the
transformation is
\begin{eqnarray}\label{gammaFAB_1}
\gamma _{FAB_{1}} =
&&\left(
  \begin{array}{cccc}
    F_{11} & F_{12} & F_{13} & F_{14} \\
    F_{21} & F_{22} & F_{23} & F_{24} \\
    F_{31} & F_{32} & V\mathbb{I} & \sqrt{T_{A}(V^{2}-1)}\sigma _{z} \\
    F_{41} & F_{42} & \sqrt{T_{A}(V^{2}-1)}\sigma _{z} & T_{A}(V+\chi_{A})\mathbb{I} \\
  \end{array}
\right),\nonumber\\
\end{eqnarray}
where $V$ is the variance of the EPR state, $T_{A}$ and $\chi_{A}$
characterize the influence of the Gaussian source noise on the
signal mode, $\mathbb{I}$ is the $2\times 2$ identity matrix,
$\sigma_{z}$ is the Pauli-z matrix, and each $F_{ij}$ represents an
unknown $2\times 2$ matrix describing either $F$ or its correlations
with $AB_1$.

With using the coherent-state protocol as an example, the
equivalence between the EB scheme and the practical PM scheme is
explained below. In the EB scheme Alice performs a heterodyne
detection on her side and gets two measurement results $P_{A}$ and
$X_{A}$. Bob's state $\rho_{B_{1}}$ would be projected into a
Gaussian state with covariance matrix $\gamma_{B_{1}}$ and mean
$d_{B_{1}}$ satisfying \cite{Garcia PHD}
\begin{eqnarray}\label{E-B}
\begin{split}
\gamma_{B_{1}} &= T_{A}(\chi_{A}+1)\mathbb{I}, \\
d_{B_{1}}=&\sqrt{\frac{2T_{A}(V-1)}{V+1}}(X_{A},-P_{A}).
\end{split}
\end{eqnarray}
In the PM scheme Alice originally prepares the signal mode $B_{0}$
in a coherent state with displacement vector
$d_{B_{1}}/{\sqrt{T_{A}}}$. Then, the effect of the source noise can
be described by
$\hat{p}_{B_{1}}=\sqrt{T_{A}}(\hat{p}_{B_{0}}+\delta_{P})$ and
$\hat{x}_{B_{1}}=\sqrt{T_{A}}(\hat{x}_{B_{0}}+\delta_{X})$, in which
$\delta_{P}$ and $\delta_{X}$ are uncorrelated noise terms with zero
mean and variance $V(\delta_{P})=V(\delta_{X})=\chi_{A}$. The state
sent to Bob is then identical to the one described in Eq.
(\ref{E-B}), indicating that source preparation in the real PM
scheme can be properly characterized using the EB scheme. As the
source-noise effect goes to zero, the equivalence would be identical
to the one described in \cite{Grosshans_QIC_03}.

When the signal is sent through the channel, the attack of the
potential eavesdropper Eve can be described as performing a unitary
transformation $U_{BE}$ over the signal mode $B_{1}$ and her modes.
After Eve's interaction, the covariance matrix of the state
$\rho_{FAB}$ would be
\begin{align}\label{gammaFAB}
&\gamma _{FAB} =\nonumber\\
&\left(
  \begin{array}{cccc}
    F_{11} & F_{12} & F_{13} & F_{14}' \\
    F_{21} & F_{22} & F_{23} & F_{24}' \\
    F_{31} & F_{32} & V\mathbb{I} & \sqrt{TT_{A}(V^{2}-1)}\sigma _{z} \\
    F_{41}' & F_{42}' & \sqrt{TT_{A}(V^{2}-1)}\sigma _{z} & T[T_{A}(V+\chi _{A})+\chi]\mathbb{I} \\
  \end{array}
\right),\nonumber\\
\end{align}
where $T$ and $\chi$ are channel parameters and $F_{ij}'$ indicates
the changed correlation terms due to Eve's interaction. Note that
since $\gamma_{FAB}$ is partly unknown but fixed, we can prove the
optimality of Gaussian attack, as shown in Appendix A. In the
following, the lower bounds on the secret key rate of this model
would be derived without knowing the exact state of Fred in both
reverse and direct reconciliations.

\section{REVERSE RECONCILIATION}

In reverse reconciliation, the secret key rate is given by
\begin{equation}\label{RR}
K_{RR}=I(a:b)-S(b:E),
\end{equation}
where $I(a:b)$ is classical mutual information between Alice and Bob
and $S(b:E)$ is quantum mutual information between Bob and Eve.
Given the above covariance matrix $\gamma_{FAB}$, $I(a:b)$ can be
calculated from the reduced matrix $\gamma _{AB}$, while $S(b:E)$
can not be learned directly from $\gamma _{FAB}$ since $F_{ij}$
contains undetermined parameters. Fortunately, another Gaussian
state $\rho_{FAB}'$ with determined covariance matrix
$\gamma_{FAB}'$ exists, and serves as an upper bound on the
calculation of the quantity $S(b:E)$. $\gamma_{FAB}'$ has the form
\begin{align}\label{gamma'FAB}
&\gamma _{FAB}' =\nonumber\\
&\left(
  \begin{array}{cccc}
    I & 0 & 0 & 0 \\
    0 & I & 0 & 0 \\
    0 & 0 & T_{A}(V+\chi _{A})\mathbb{I} & \sqrt{T[T_{A}^{2}(V+\chi _{A})^{2}-1]}\sigma _{z} \\
    0 & 0 & \sqrt{T[T_{A}^{2}(V+\chi _{A})^{2}-1]}\sigma _{z} & T[T_{A}(V+\chi _{A})+\chi]\mathbb{I} \\
  \end{array}
\right).\nonumber\\
\end{align}
The relationship between two Gaussian states with $\gamma_{FAB}$ and
$\gamma_{FAB}'$ is explained below. Considering the pure Gaussian
state $\rho'_{FAB_{1}}$ with the covariance matrix
\begin{eqnarray}\label{gamma'FAB1}
 &&\gamma _{FAB_{1}}' =\nonumber \\
 &&\left(
  \begin{array}{cccc}
    I & 0 & 0 & 0 \\
    0 & I & 0 & 0 \\
    0 & 0 & T_{A}(V+\chi _{A})\mathbb{I} & \sqrt{T_{A}^{2}(V+\chi _{A})^{2}-1}\sigma _{z} \\
    0 & 0 & \sqrt{T_{A}^{2}(V+\chi _{A})^{2}-1}\sigma _{z} & T_{A}(V+\chi _{A})\mathbb{I} \\
  \end{array}
\right).\nonumber\\
\end{eqnarray}
The reduced state $\rho'_{B_{1}}=\mathrm{Tr}_{FA}(\rho'_{FAB_1})$ is
identical to the reduced state
$\rho_{B_{1}}=\mathrm{Tr}_{FA}(\rho_{FAB_1})$, so $\rho'_{FAB_1}$
and $\rho_{FAB_1}$ are two different purifications of this state.
According to \cite{Nielsen_Chuang_BOOK_00}, one purification of a
fixed system can be transformed into another through a local unitary
transformation on its ancillary system. Hence there exists such a
unitary map $U_{FA}$ that transforms $\rho_{FAB_1}$ to
$\rho'_{FAB_1}$. Furthermore, after taking Eve's attack $U_{BE}$
into account and noticing that $U_{FA}$ and $U_{BE}$ commute, it can
be proved that $\rho_{FAB}$ will be transformed into $\rho'_{FAB}$
through $U_{FA}$.

In the rest of the paper expressions with the prime indicate the
terms calculated by $\gamma'_{FAB}$. The following lemma then allows
us to bound Eve's knowledge.

\textit{Lemma 1.}  \emph{Given two Gaussian states $\rho_{FAB}$ and
$\rho'_{FAB}$ with covariance matrices $\gamma_{FAB}$ and
$\gamma'_{FAB}$ shown in Eqs. (\ref{gammaFAB}) and
(\ref{gamma'FAB}), respectively, one has the equality}
\begin{equation}\label{bE=bE'}
S(b:E) = S'(b:E).
\end{equation}

\emph{Proof.} Based on $\gamma_{FAB}$ and $\gamma'_{FAB}$ the mutual
information between Bob and Eve is, respectively, given as
\begin{equation}\label{comprev}
\begin{split}
S(b:E)&= S(E)-S(E\mid b), \\
S'(b:E)&= S'(E)-S'(E\mid b),
\end{split}
\end{equation}
where $S(E)$ and $S'(E)$ are the von Neumann entropy of Eve's state,
and $S(E\mid b)$ and $S'(E\mid b)$ are Eve's entropy conditioned on
Bob's measurement results. $S(E)=S(F,A,B)$ and $S'(E)=S'(F,A,B)$ can
be verified from the fact that Eve could purify the Fred-Alice-Bob
system \cite{Garcia PHD}. Because $\rho_{FAB}$ can be changed into
$\rho'_{FAB}$ through a unitary transformation $U_{FA}$, the von
Neumann entropy $S(F,A,B)=S'(F,A,B)$, and thus $S(E)=S'(E)$. On the
other hand, conditioning on Bob's result $b$, the conditional state
with $\gamma_{FA|B=b}$ can be transformed into the one with
${\gamma'}_{FA|B=b}$ through $U_{FA}$, and thus $S(F,A\mid
b)=S'(F,A\mid b)$. Combining another fact that $S(E\mid b)=S(F,A\mid
b)$ and $S'(E\mid b)=S'(F,A\mid b)$, we conclude that
$S(b:E)=S'(b:E)$. \rightline{$\Box$}

Lemma 1 implies that calculation with $\gamma_{FAB}'$ can bound
Eve's knowledge. Note that Eq. (\ref{bE=bE'}) is valid for protocols
implementing either squeezed-state or coherent-state protocol with
Bob using homodyne or heterodyne detection. Hence our model provides
a tight security bound for all these protocols in reverse
reconciliation.

\section{DIRECT RECONCILIATION}
Though direct reconciliation has the 3dB limit, the security bounds
for the sqeezed-state protocol with homodyne detection and the
no-switching protocol \cite{Weedbrook_PRL_04} are analyzed
theoretically. In direct reconciliation, the secret key rate is
given by
\begin{equation}\label{DR}
K_{DR}=I(a:b)-S(a:E).
\end{equation}
$I(a:b)$ can be calculated from $\gamma_{AB}$, and $S(a:E)$ can be
bounded by the following lemma.

\textit{Lemma 2.} \emph{Given two Gaussian states $\rho_{FAB}$ and
$\rho'_{FAB}$ with covariance matrices $\gamma_{FAB}$ and
$\gamma'_{FAB}$, the following inequality can be verified}
\begin{equation}\label{aE<aE'}
S(a:E) \leq S'(a:E).
\end{equation}
The proof can be seen in Appendix B. Note that the equality in Eq.
(\ref{aE<aE'}) is achieved only when $F$ is independent of $E$,
which is not necessarily satisfied in practice. This means that in
order to bound the secret key rate, Eve's knowledge about Alice is
overestimated by using $S'(a:E)$. Thus, the security bound derived
here is not tight.

\section{NUMERICAL SIMULATION}

Our simulation concerns the no-switching protocol in both reverse
and direct reconciliations. The secret key rate $K_{DR}$ or $K_{RR}$
would depend on the variables $V$, $T_{A}$, $\chi _{A}$, $T$ and
$\chi$ characterizing either source or channel influences. In the
simulation the variance is set to $V=20$ and channel excess noise
$\epsilon=T\chi-1+T=0.04$ close to the practical scenario
\cite{Lodewick_Ex_PRA_07}, where electronic noise in Bob's detection
is simply treated as part of $\epsilon$. In addition, to analyze
both signal attenuation and amplification cases the source
parameters are set to $\epsilon _{A}=T_{A}\chi_{A}-1+T_{A}=0.1$ and
$T_{A}=0.9$ or $T_{A}=1.1$ with regard to each process.

The secret key rate is calculated using our model, the untrusted
source noise model, and the beam-splitter model. The mutual
information $I(a:b)$ is calculated according to the protocol used,
whose formula can be found in \cite{Garcia PHD}. $S(a:E)$ and
$S(b:E)$ in our model can be bounded using the simplified covariance
matrix $\gamma'_{FAB}$. To deal with the untrusted source noise,
Fred is assumed to be part of Eve, and thus $S(a:E)$ and $S(b:E)$
are derived from $\gamma_{AB}$ \cite{Shen_JPB_09}. For the
beam-splitter model the key rate is calculated with the covariance
matrix including the ancillary modes, which is given in Eqs.
(\ref{gamma FGAB<1}) and (\ref{gamma FGAB>1}) in Appendix C.

\begin{figure}[t]
\includegraphics[width=9.0 cm]{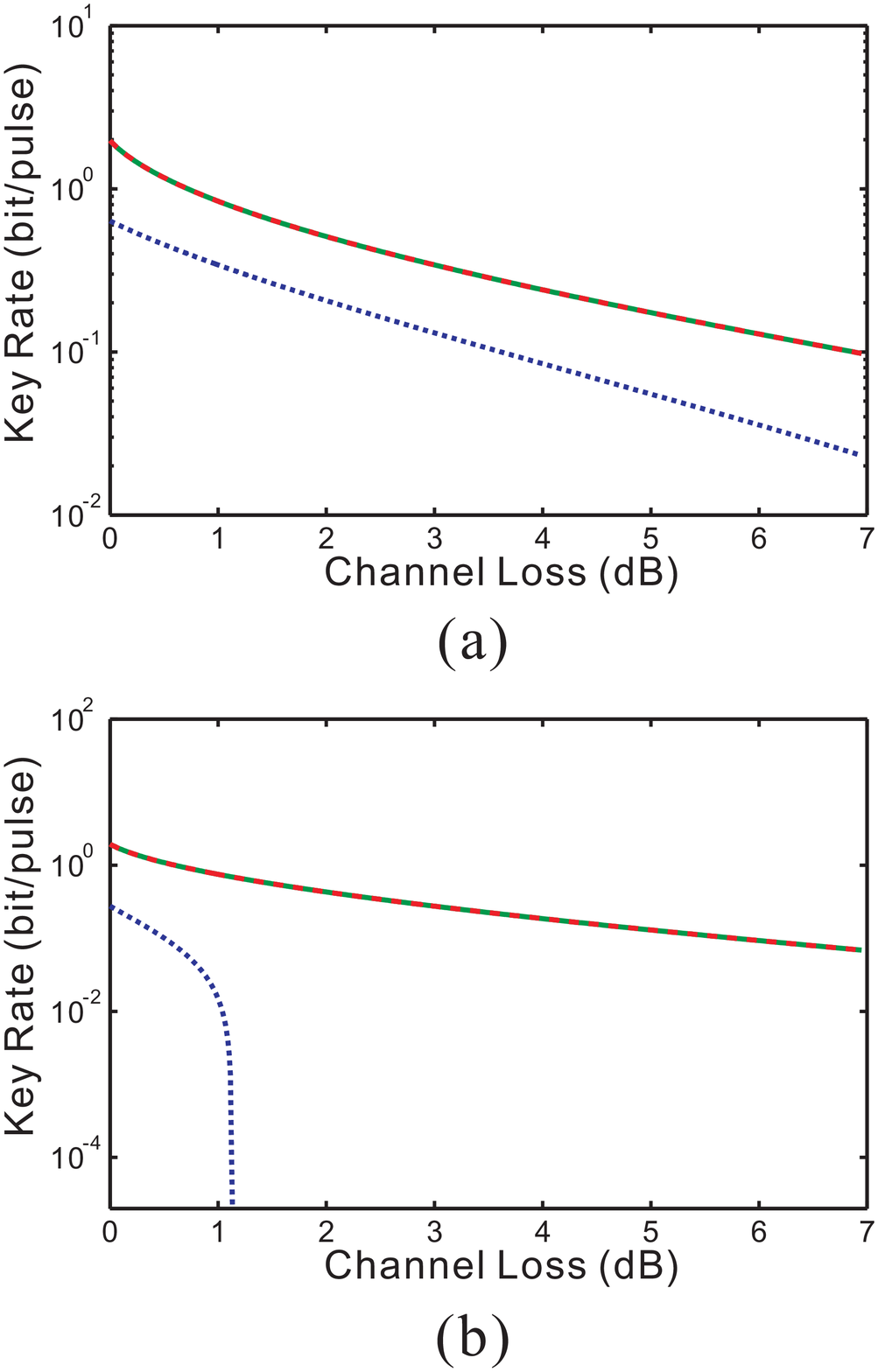}
\caption{Secret key rate as a function of the transmittance of the
channel in no-switching protocol with reverse reconciliation. The
solid line stands for our model, dashed line for the beam-splitter
model, and dotted line for the untrusted source noise model. Data
are acquired under the variance of $V=20$, and channel's excess
noise is chosen to be $\epsilon=0.04$. (a) Signal attenuation with
source parameter $T_{A}=0.9$ and $\epsilon _{A}=0.1$. (b) Signal
amplification with source parameter $T_{A}=1.1$ and $\epsilon
_{A}=0.1$}.\label{Fig-2}
\end{figure}

\begin{figure}[t]
\includegraphics[width=9.0 cm]{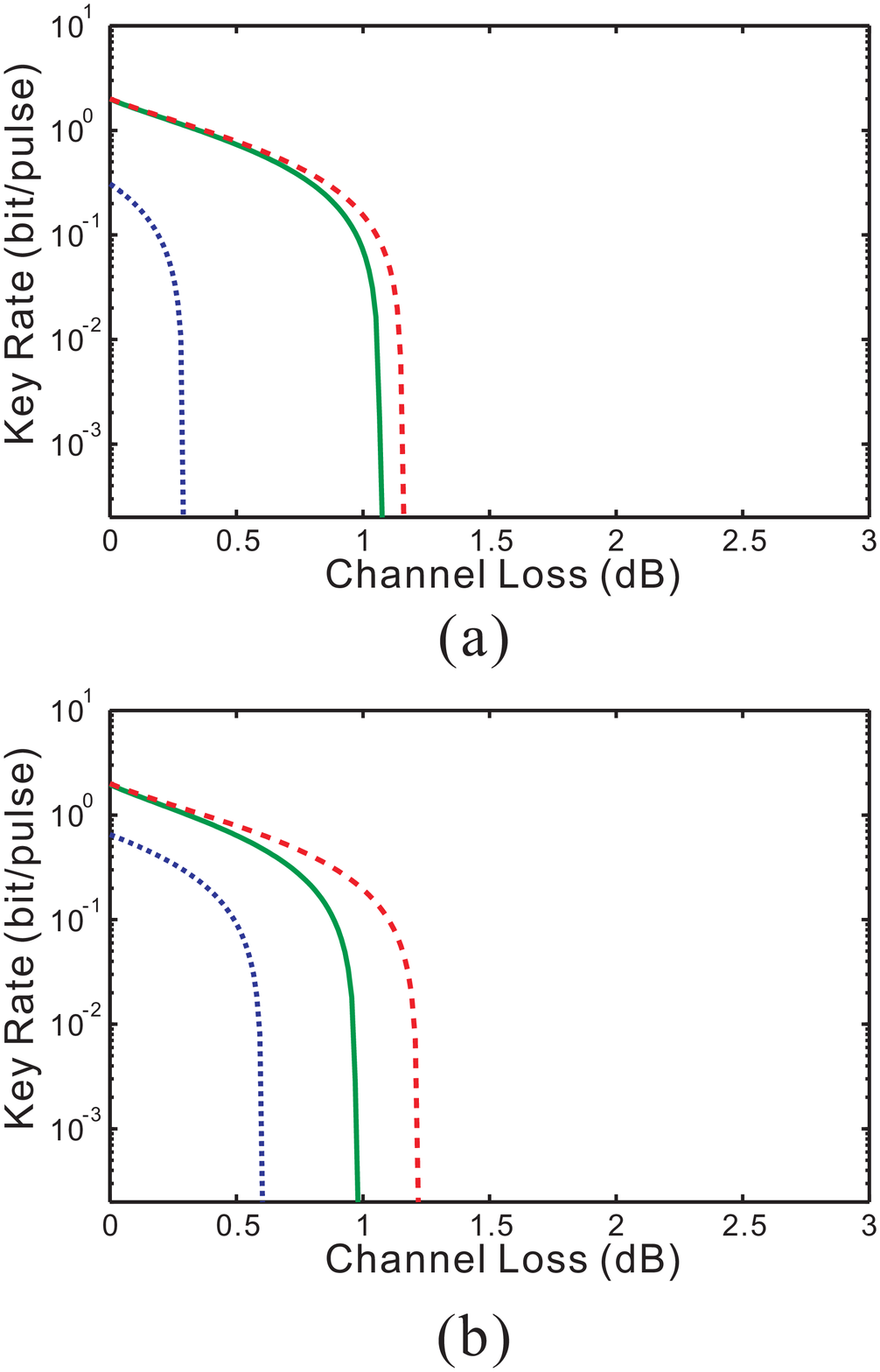}
\caption{Secret key rate as a function of the transmittance of the
channel in no-switching protocol with direct reconciliation. The
solid line stands for our model, dashed line for the beam-splitter
model, and dotted line for the untrusted source noise model. Data
are acquired under the variance of $V=20$, and channel's excess
noise is chosen to be $\epsilon=0.04$. (a) signal attenuation with
source parameter $T_{A}=0.9$ and $\epsilon _{A}=0.1$. (b) signal
amplification with source parameter $T_{A}=1.1$ and $\epsilon
_{A}=0.1$}.\label{Fig-3}
\end{figure}

\section{DISCUSSION AND CONCLUSION}

The simulation results can be seen in Figs. \ref{Fig-2} and
\ref{Fig-3}, where performances of our model, the beam-splitter
model, and the untrusted source noise model are shown under
no-switching protocol in reverse and direct reconciliations. From
Figs. \ref{Fig-2} and \ref{Fig-3}, it is clearly seen that the
secret key rate of our model (solid line) coincides with that of the
beam-splitter model (dashed line) in reverse reconciliation, while
in direct reconciliation our result is lower. In addition, the
security bound of our model is significantly higher than the
untrusted source noise model (dotted line) in all cases.


In the reverse reconciliation case (as shown in Fig. \ref{Fig-2})
the coincidence of our model and the beam-splitter model on the
secret key rate means that our model provides a tight security
bound, even by generalizing Fred's interaction. This coincidence is
due to the fact that both models provide the same signal state, and
the information leakage to Eve is estimated through this state. In
the direct reconciliation shown in Fig. \ref{Fig-3}, we remark that
our bound on the secret key rate can be further improved since the
information gained by Eve is overestimated in mathematical
treatment.

In conclusion, we have proposed a model to characterize the general
Gaussian source noise in CV QKD.  The result coincides with that of
the beam-splitter model in reverse reconciliation protocols, proving
that our generalized model can provide a tight bound on the secret
key rate. In direct reconciliation, though the security bound is not
tight, it still surpasses that of the untrusted source noise model
in a significant way.

\section*{Acknowledgments}
This work is supported by the Key Project of National Natural
Science Foundation of China (Grant No. 60837004), National Hi-Tech
Research and Development (863) Program and the Xiao Zhang Foundation
of Peking University. X. Peng acknowledges the support from China
Postdoctoral Science Foundation funded project. The authors also
thank R. Filip's group for fruitful discussions.

\begin{appendix}

\section{Optimality of Gaussian Attack}

Here, in our model, the security analysis on the optimality of
Gaussian collective attack needs to be rechecked because the neutral
party Fred is introduced in the EB scheme. Given the state
$\rho_{FAB}$ with covariance matrix $\gamma_{FAB}$, it can be
demonstrated that the security bound is obtained by considering
Gaussian attack. The reason is listed as follows.

In Eqs. (\ref{RR}) and (\ref{DR}) $I(a:b)$ is lower bounded by
Gaussian attack \cite{Navascues_PRL_06}. As for $S(a:E)$ in direct
reconciliation, considering Bob and Fred together as a larger state
$B^{*}$, thus $S(a:E)=S(E)-S(E\mid a)=S(A,B^{*})-S(B^{*}\mid a)$.
According to \cite{Navascues_PRL_06}, $S(a:E)$ reaches its maximum
when the quantum state $\rho_{AB^{*}}$ or $\rho_{FAB}$ is Gaussian,
with covariance matrix $\gamma _{FAB}$. Therefore, Gaussian attack
is optimal for direct reconciliation protocols. The deduction for
the reverse reconciliation protocols follows a similar route.

Note that, in practical PM scheme, the source noise including light
intensity fluctuation from a laser or modulator is a Gaussian one,
thus the analysis is limited to the situation where Fred performs a
general Gaussian transformation.

\section{Proof of Ineq. (11)}


The mutual information between Alice and Eve is given by
\begin{equation}\label{compdir}
\begin{split}
S(a:E)&= S(E)-S(E\mid a), \\
S'(a:E)&= S'(E)-S'(E\mid a),
\end{split}
\end{equation}
where the equation $S(E)=S'(E)$ can be verified with similar reason
demonstrated in Sec. {III}. Furthermore, we use the relations
\begin{eqnarray}
S(E\mid a) &\geq& S(E\mid a,f), \label{E|a>E|af}\\
S(E\mid a,f) &\geq& S'(E\mid a,f), \label{E|af>'E|af}\\
S'(E\mid a,f) &=& S'(E\mid a), \label{'E|af='E|a}
\end{eqnarray}
where $S(E\mid a,f)$ ($S'(E\mid a,f)$) means Eve's entropy
conditioned on the measurement results $a$ ($a'$) and $f$ ($f'$) of
Alice and Fred. Here, Eq. (\ref{'E|af='E|a}) is obtained by noticing
that in Eq. (\ref{gamma'FAB}) Fred is uncorrelated with the rest of
the system, that is, $\rho'_{FABE}=\rho'_{F}\bigotimes\rho'_{ABE}$,
so \cite{Nielsen_Chuang_BOOK_00}
\begin{eqnarray}
S'(E\mid a,f)&=&S'(E; a, f)-S'(a,f)\nonumber\\
&=&S'(E; a)+S'(f)-[S'(a)+S'(f)]\nonumber \\
&=&S'(E\mid a).
\end{eqnarray}
On the other hand, Eq. (\ref{E|a>E|af}) holds because of the strong
subadditivity of the von Neumann entropy
\cite{Nielsen_Chuang_BOOK_00}. Furthermore, the equation
\begin{equation}\label{Eaf=Eaf'}
S(E\mid a,f)=S'(E\mid a,f),
\end{equation}
can be verified in the squeezed-state protocol with homodyne
detection, while
\begin{equation}\label{Eaf>Eaf'}
S(E\mid a,f) \geq S'(E\mid a,f),
\end{equation}
holds in the no-switching protocol. Because both Gaussian states
$\rho_{FABE} $ and $\rho'_{FABE}$ are pure states, one has
\begin{equation}\label{Eaf=Baf}
\begin{split}
S(E\mid a,f)&=S(B\mid a,f),\\
S'(E\mid a,f)&=S'(B\mid a,f).
\end{split}
\end{equation}
With using Eq. (\ref{Eaf=Baf}), Eq. (\ref{Eaf=Eaf'}) and InEq.
(\ref{Eaf>Eaf'}) are explained below.

  (1) In the squeezed-state protocol with homodyne detection, $S(B\mid
a,f)= S'(B\mid a,f)$ can be verified through proving $\gamma
_{B}^{af} = {\gamma'}_{B}^{af}$, in which $\gamma_{B}^{af}$
(${\gamma'}_{B}^{af}$) means the covariance matrix of the state
$\rho_B$ ($\rho'_B$) conditioning on the measurement results $a$
($a'$) and $f$ ($f'$) of the state $\rho_{FA}$ ($\rho'_{FA}$). The
covariance matrix $\gamma _{B}^{af}$ can be obtained by \cite{Garcia
PHD}
\begin{equation}\label{gammaBaf}
\gamma _{B}^{af}= \gamma _{B}- \sigma _{B-FA} (X\gamma _{FA}X)^{MP}
\sigma _{B-FA}^{T},
\end{equation}
where $\sigma _{B-FA}$, $\gamma _{FA}$ and $\gamma _{B}$ denote part
of $\gamma _{FAB}$
\begin{equation}\label{constgammaFAB}
\gamma_{FAB}=\left(
  \begin{array}{cc}
    \gamma _{FA} & \sigma _{B-FA}^{T} \\
    \sigma _{B-FA} & \gamma _{B} \\
  \end{array}
\right),
\end{equation}
and $X$ is a matrix of the form
\begin{eqnarray}\label{XX}
X = \left(
  \begin{array}{cccccc}
    1 & 0 & 0 & 0 & 0 & 0 \\
    0 & 0 & 0 & 0 & 0 & 0 \\
    0 & 0 & 1 & 0 & 0 & 0 \\
    0 & 0 & 0 & 0 & 0 & 0 \\
    0 & 0 & 0 & 0 & 1 & 0 \\
    0 & 0 & 0 & 0 & 0 & 0 \\
  \end{array}
\right),
\end{eqnarray}
which stands for the homodyne detection process on the $x$
quadrature. Note that the situation where the $p$ quadrature is
measured has been omitted as its analysis would be identical to the
$x$ quadrature case. The unitary transformation $U_{FA}$ corresponds
in phase-space to a symplectic operation $S$ \cite{Garcia PHD}, and
therefore $\gamma_{FAB}'=(S\bigoplus
\mathbb{I}_{B})\gamma_{FAB}(S\bigoplus \mathbb{I}_{B})^{T}$.
Combining Eq. (\ref{gammaBaf}), ${\gamma'}_{B}^{af}$ would then be
\begin{equation}\label{gammaBaf'}
{\gamma'}_{B}^{af} = \gamma _{B}- \sigma _{B-FA}S^{T}(XS\gamma
_{FA}S^{T}X)^{MP}S\sigma _{B-AF}^{T}.
\end{equation}
Without loss of generality, we assume $S$ takes a general form
\cite{Garcia PHD}
\begin{equation}\label{S}
S = \left(
    \begin{array}{cccccc}
      a & 0 & b & 0 & c & 0 \\
      0 & a' & 0 & b' & 0 & c' \\
      d & 0 & e & 0 & f & 0 \\
      0 & d' & 0 & e' & 0 & f' \\
      g & 0 & h & 0 & i & 0 \\
      0 & g' & 0 & h' & 0 & i' \\
    \end{array}
  \right),
\end{equation}
and hence satisfying $SX=XS$. Therefore, one has
\begin{eqnarray}\label{simplify}
(XS\gamma _{AF}S^{T}X)^{MP}&=&(SX\gamma _{AF}XS^{T})^{MP}\nonumber \\
&=&(S^{T})^{-1}(X\gamma _{AF}X)^{MP}(S)^{-1},
\end{eqnarray}
according to the characteristics of the Moore-Penrose pseudoinverse
of matrix \cite{Isreal_Greville_BOOK_03}. Observing Eqs.
(\ref{gammaBaf}), (\ref{gammaBaf'}) and (\ref{simplify}), $\gamma
_{B}^{af} = {\gamma'} _{B}^{af}$, which means $S(B\mid a,f)=
S'(B\mid a,f)$. According to Eq. (\ref{Eaf=Baf}), the validity of
Eq. (\ref{Eaf=Eaf'}) is proved.

  (2) In the no-switching protocol, InEq. (\ref{Eaf>Eaf'}) is verified
by comparing the explicit von Neumann entropy calculated from the
two covariance matrices ${\gamma'}_{B}^{af}$ and $\gamma_{B}^{af}$.
Starting from $\rho'_{FAB}$ , ${\gamma'}_{B}^{af}$ can be written as
\cite{Garcia PHD}
\begin{eqnarray}\label{gammaBaf'2}
{\gamma'}_{B}^{af} &=&\gamma _{B}'- \sigma _{B-FA}'(\gamma
_{FA}'+\mathbb{I})^{-1}\sigma'^{T}_{B-FA}\nonumber \\
&=&\{T[T_{A}(V+\chi_{A})+\chi]-\frac{T[T_{A}^{2}(V+\chi_{A})^{2}-1]}{T_{A}(V+\chi_{A})+1}\}\mathbb{I},\nonumber\\
\end{eqnarray}
where $\mathbb{I}$ is the 2$\times$2 identity matrix. On the other
hand,
\begin{eqnarray}\label{gammaBaf2}
\gamma _{B}^{af} =\gamma _{B}'-\sigma
_{B-FA}'(S^{-1})^{T}[S^{-1}\gamma
_{FA}'(S^{-1})^{T}\nonumber+\mathbb{I}]^{-1}S^{-1}\sigma'^{T}_{B-FA}.\nonumber\\
\end{eqnarray}
For the symplectic transformation $S^{-1}$, a decomposition of the
form $S^{-1} = PS_{r}Q$ exists, which is known as the Bloch-Messiah
reduction \cite{BM reduction}. Here, $S_{r}$ is a squeezing operator
on each mode
\begin{equation}\label{Sr}
S_{r} = \left(
  \begin{array}{cccccc}
    e^{s_{1}} & 0 & 0 & 0 & 0 & 0 \\
    0 & e^{-s_{1}} & 0 & 0 & 0 & 0 \\
    0 & 0 & e^{s_{2}} & 0 & 0 & 0 \\
    0 & 0 & 0 & e^{-s_{2}} & 0 & 0 \\
    0 & 0 & 0 & 0 & e^{s_{3}} & 0 \\
    0 & 0 & 0 & 0 & 0 & e^{-s_{3}} \\
  \end{array}
\right),
\end{equation}
and $P$ and $Q$ stand for two passive transformations satisfying
$P^{T}P=\mathbb{I}$ and $Q^{T}Q=\mathbb{I}$. Without loss of
generality, matrix $Q$ takes a general form
\begin{equation}\label{Q}
Q = \left(
    \begin{array}{cccccc}
      a & 0 & b & 0 & c & 0 \\
      0 & a' & 0 & b' & 0 & c' \\
      d & 0 & e & 0 & f & 0 \\
      0 & d' & 0 & e' & 0 & f' \\
      g & 0 & h & 0 & i & 0 \\
      0 & g' & 0 & h' & 0 & i' \\
    \end{array}
  \right).
\end{equation}
With implementing the orthogonality of the passive transformation
$P$, one has
\begin{eqnarray}\label{simplify2}
\gamma _{B}^{af} &=&\gamma _{B}'-\sigma
_{B-FA}'Q^{T}{S_{r}}^{T}(S_{r}Q\gamma
_{FA}'Q^{T}{S_{r}}^{T}+\mathbb{I})^{-1}S_{r}Q{\sigma _{B-FA}'}^{T}\nonumber \\
&=& T[T_{A}(V+\chi_{A})+\chi]\mathbb{I}-\left(
  \begin{array}{cc}
    \frac{T[T_{A}^{2}(V+\chi_{A})^{2}-1]}{T_{A}(V+\chi_{A})-1+\frac{1}{W}} & 0 \\
    0 & \frac{T[T_{A}^{2}(V+\chi_{A})^{2}-1]}{T_{A}(V+\chi_{A})-1+\frac{1}{1-W}} \\
  \end{array}
\right),\nonumber\\
\end{eqnarray}
where letter $W$ represents
\begin{equation}\label{W}
\begin{split}
W &= \frac{e^{2s_{1}}c^{2}}{e^{2s_{1}}+1} + \frac{e^{2s_{2}}f^{2}}{e^{2s_{2}}+1} + \frac{e^{2s_{3}}i^{2}}{e^{2s_{3}}+1},\\
1-W &= \frac{c^{2}}{e^{2s_{1}}+1} + \frac{f^{2}}{e^{2s_{2}}+1} +
\frac{i^{2}}{e^{2s_{3}}+1}.
\end{split}
\end{equation}
Using $Q^{T}Q=\mathbb{}{I}$, $W$ takes its value within $0 < W < 1$.
To calculate its entropy, note that the von Neumann entropy of a
Gaussian state $\rho$ is given by
\begin{equation}
S(\rho) = \sum_{i} g( \frac{\lambda _{i} -1}{2}),
\end{equation}
where $g(x) = (x+1)\log_{2}(x+1) - x\log_{2}(x)$ and $\lambda _{i}$
is the symplectic eigenvalue of the covariance matrix of $\rho$. It
can then be shown that the von Neumann entropy of $\gamma _{B}^{af}$
increases as its symplectic eigenvalue increases. Furthermore, the
symplectic eigenvalue of $\gamma _{B}^{af}$ is the square of the
multiplication of its diagonal entries, and the minimum of this
eigenvalue is reached when $W=\frac{1}{2}$ in
$\gamma_{B}^{af}={\gamma'}_{F}^{af}$. This yields InEq.
(\ref{Eaf>Eaf'}) in the no-switching protocol.

$ $

$ $

\section{Beam-Splitter Model under Gaussian Channel}


The secret key rate of the beam-splitter model is to be calculated
with signal attenuation or amplification, respectively. In case of
attenuation ($T_{A}<1$) the model is shown in Fig. 1(a)
\cite{Filip_PRA_10} and the result is obtained by setting the
parameters in Eq. (\ref{gammaFAB}) as
\begin{widetext}
\begin{equation}\label{gamma FGAB<1}
\gamma _{FGAB} =\left(
  \begin{array}{cccc}
    N & \sqrt{T_{A}(N^{2}-1)}\sigma_{z} & 0 & -\sqrt{T(1-T_{A})(N^{2}-1)}\sigma_{z} \\
    \sqrt{T_{A}(N^{2}-1)}\sigma_{z} & [T_{A}N+(1-T_{A})V]\mathbb{I} & \sqrt{(1-T_{A})(V^{2}-1)}\sigma_{z} & \sqrt{TT_{A}(1-T_{A})}(V-N)\mathbb{I} \\
    0 & \sqrt{(1-T_{A})(V^{2}-1)}\sigma_{z} & V\mathbb{I} & \sqrt{TT_{A}(V^{2}-1)}\sigma _{z} \\
    -\sqrt{T(1-T_{A})(N^{2}-1)}\sigma_{z} & \sqrt{TT_{A}(1-T_{A})}(V-N)\mathbb{I} & \sqrt{TT_{A}(V^{2}-1)}\sigma _{z} & T[T_{A}(V+\chi _{A})+\chi]\mathbb{I} \\
  \end{array}
\right),
\end{equation}
where $N$ is the variance of the ancillary EPR' shown in Fig. 1(a),
which is related to the source parameters through
$N=T_{A}\chi_{A}/(1-T_{A})$. Using this specific form of
$\gamma_{FAB}$, Eve's knowledge $S(E)-S(E\mid a)$ can be calculated
by implementing the relation
\begin{equation}
S(E)-S(E\mid a)=S(FGAB)-S(FGB\mid a).
\end{equation}

In case of amplification ($T_{A}>1$), one needs to modify the
parameter setting, and change the model according to Fig. 1(b).
Under this situation, the global covariance matrix reads
\begin{equation} \label{gamma FGAB>1}
\gamma _{FGAB}=\left(
  \begin{array}{cccc}
    N_{B} & \sqrt{T_{B}(N^{2}_{B}-1)}\sigma_{z} &  -\sqrt{(1-T_{B})(N^{2}_{B}-1)}\sigma_{z} & 0 \\
    \sqrt{T_{B}(N^{2}_{B}-1)}\sigma_{z} & [T_{B}N_{B}+(1-T_{B})V_{B}]\mathbb{I} & \sqrt{T_{B}(1-T_{B})}(V_{B}-N_{B})\mathbb{I} & \sqrt{T(1-T_{B})(V^{2}_{B}-1)}\sigma_{z}\\
    -\sqrt{(1-T_{B})(N^{2}_{B}-1)}\sigma_{z} & \sqrt{T_{B}(1-T_{B})}(V_{B}-N_{B})\mathbb{I} & T_{B}(V_{B}+\chi_{B})\mathbb{I} & \sqrt{TT_{B}(V^{2}_{B}-1)}\sigma _{z} \\
    0  & \sqrt{T(1-T_{B})(V^{2}_{B}-1)}\sigma_{z} & \sqrt{TT_{B}(V^{2}_{B}-1)}\sigma _{z} &  T(V_{B}+\chi)\mathbb{I}\\
  \end{array}
\right),
\end{equation}
\end{widetext}
where $V_{B}=T_{A}(V+\chi_{A})$ is the modified variance of the EPR
state, and the corresponding noise parameters are
$T_{B}=T_{A}(V^2-1)/[T_{A}^2(V+\chi_{A})^2-1]$ and
$\chi_{B}=[T_{A}^2(V+\chi_{A})(V\chi_{A}+1)-V]/[T_{A}(V^2-1)]$,
leading to a modified variance of the ancillary EPR' reading
$N_{B}=T_{B}\chi_{B}/(1-T_{B})$. It is easy to verify that such
replacement would lead to the same $\gamma_{AB}$ as in Eq.
(\ref{gammaFAB})
\begin{eqnarray}
\gamma_{AB}=&&\left(
  \begin{array}{cc}
    T_{B}(V_{B}+\chi_{B})\mathbb{I} & \sqrt{TT_{B}(V^{2}_{B}-1)}\sigma _{z} \\
    \sqrt{TT_{B}(V^{2}_{B}-1)}\sigma _{z} &  T(V_{B}+\chi)\mathbb{I}\\
  \end{array}
\right)\nonumber\\ = &&\left(
           \begin{array}{cc}
             V\mathbb{I} & \sqrt{TT_{A}(V^{2}-1)}\sigma _{z} \\
             \sqrt{TT_{A}(V^{2}-1)}\sigma _{z} & T[T_{A}(V+\chi _{A})+\chi]\mathbb{I} \\
           \end{array}
         \right),
\end{eqnarray}
and is therefore able to describe the amplification process. In
order to make the model physical realizable, the parameters also
need to satisfy $T_{B}<1$ and $\chi_{B}\geq(1-T_{B})/T_{B}$. The
first inequality is easily recognized since now
$T_{A}^2(V+\chi_{A})^2>T_{A}V^2$ and $T_{A}>1$, leading to
$T_{A}(V^2-1)<T_{A}^2(V+\chi_{A})^2-1$. For the second inequality,
by substituting $T_{A}$, $\chi_{A}$ and $V$ into it, we can
transform it into
\begin{equation}
\chi_{A}^{2}+(V-1)\chi_{A}-\frac{(T_{A}V-1)(T_{A}-1)}{T_{A}^{2}}\geq
0.
\end{equation}
Given that $\chi_{A}$ satisfies $\chi_{A}\geq (T_{A}-1)/T_{A}$ when
$T_{A}>1$, it is easy to verify that the left hand side reaches its
minimum when $\chi_{A}=(T_{A}-1)/T_{A}$, and the minimum is just 0,
which proves the inequality.

With the above covariance matrix Eq. (\ref{gamma FGAB>1}), Eve's
knowledge can be obtained.

\end{appendix}

\end{document}